# Modelling Realistic TiO$_2$ Nanospheres:
# a Benchmark Study of SCC-DFTB against Hybrid DFT

Daniele Selli, Gianluca Fazio and Cristiana Di Valentin[*]

Dipartimento di Scienza dei Materiali, Università di Milano-Bicocca,
via Cozzi 55 20125, Milano, Italy

**ABSTRACT:** TiO$_2$ nanoparticles (NPs) are nowadays considered fundamental building blocks for many technological applications. Morphology is found to play a key role with spherical NPs presenting higher binding properties and chemical activity. From the experimental point of view, the characterization of these nano-objects is extremely complex, opening a large room for computational investigations. In this work, TiO$_2$ spherical NPs of different size (from 300 to 4000 atoms) have been studied with a two-scale computational approach. Global optimization to obtain stable and equilibrated NSs was performed with a self-consistent charge density functional tight-binding (SCC-DFTB) simulated annealing process, causing a considerable atomic rearrangement within the nanospheres. Those SCC-DFTB relaxed structures have been then optimized at DFT(B3LYP) level of theory. We present a systematic and comparative SCC-DFTB vs DFT(B3LYP) study of the structural properties, with particular emphasis on the surface-to-bulk sites ratio, coordination distribution of surface sites and surface energy. From the electronic point of view, we compare HOMO–LUMO and Kohn-Sham gaps, total and projected density of states. Overall, the comparisons between DFTB and hybrid DFT show that DFTB provides a rather accurate geometrical and electronic description of these nanospheres of realistic size (up to a diameter of 4.4 nm) at an extremely reduced computational cost. This opens for new challenges in simulations of very large systems and more extended molecular dynamics.

[*] Corresponding author: cristiana.divalentin@mater.unimib.it

1. INTRODUCTION

TiO$_2$ nanoparticles (NPs) have been attracting increasing interest during the last decade for their outstanding and versatile physical and chemical properties. Titanium oxide NPs are used as building blocks for DSSCs materials with enhanced absorption,[1,2,3] and as materials for photocatalysis and photoelectrochemistry.[4,5,6,7] More recently TiO$_2$ NPs are gaining attention for their potential as materials for innovative biomedical applications, e.g. as nanocarriers for selective drug delivery, for imaging, sonodynamic or photodynamic therapy of cancer.[8,9,10,11]

Several methods, including sol-gel, reverse micelle, sonochemical, hydrothermal and the microwave approach, have been exploited to synthesize TiO$_2$ nanoparticles of different size and shape. Since the technological interest for TiO$_2$ NPs is strongly related to their morphology, many efforts have been done in order to tailor it by controlling the experimental conditions of preparation and by using ad hoc surface chemistry.[12,13,14,15,16,17,18]

Among all the synthetized titania-based nano-objects, high curvature nano-systems, e.g. nanospheres (NSs) or nanorods, are of prominent interest for TiO$_2$ functionalization in photoapplications and in biomedicine.[7,8] Regarding TiO$_2$ NSs, it has been demonstrated that the most interesting size range of diameter is below 20 nm, when the surface-to-bulk aspect ratio starts to approach one and the anatase phase becomes more stable than rutile.[8,19,20] Anatase phase is considered to be particularly photoactive with respect to the other stable TiO$_2$ phases.[21,22] Spherical NPs of such small size are characterized by very high surface curvature and by very high density of undercoordinated surface atoms, especially highly reactive 4-fold coordinated Ti atoms.[7,8] Those are excellent binding sites for organic and bio-molecules to form inorganic/organic nanoconjugates that combine the physico-chemical properties of the inorganic material to those of the attached molecule.

The surface complexity of highly curved TiO$_2$ nanospheres makes their experimental characterization extremely difficult, especially when one aims at an atomistic resolution. In this scenario, quantum chemical simulations are an extremely precious tool for the investigation and rationalization of the nanoparticles surface topology and of its relation with the observed physico-chemical properties.

In this work we present a systematic study to show that density functional tight-binding theory (DFTB), a less computational expensive DFT-based method, can be a reliable tool to accurately model structural and electronic properties of nanoparticles of realistic size, especially when those are not affordable at the DFT level. In particular, here we use the self-consistent charge extension of the DFTB method (SCC-DFTB), that at a reduced computational cost maintains quantum insight with an accuracy comparable to standard DFT.[23,24]

In the past years many studies have shown that SCC-DFTB can be successfully exploited to investigate semiconductors like Si/SiO$_2$ interfaces in MOSFETs,[25] hybrid inorganic-organic systems (gold-thiolate compounds,[26] organic molecules interacting with GaAs,[27] etc.), low dimensional materials like carbon,[28] MoS$_2$,[29] chrysotile[30] nanotubes or carbon[31] and MoS$_2$[32] fullerenes, transition metal and relative metalorganic complexes[33] and so on.

In the case of TiO$_2$ materials, the SCC-DFTB approach has been used to accurately calculate properties of periodic systems, such as bulk TiO$_2$ and TiO$_2$ surfaces, of both anatase and rutile phases, as well as small TiO$_2$ clusters, giving results in good agreement with ab initio (DFT) references.[34,35,36,37] Furthermore, benchmarks of the method have been performed recently also for more complex systems, like amorphous TiO$_2$[38,39] and molecule-titania surface interfaces.[40,41]

This long list of success in many fields of technological relevance validates the use of SCC-DFTB as a reliable tool for quantum investigations of extended systems.

In this work, we have considered TiO$_2$ nanospheres with a realistic diameter size between 1.5 and 4.4 nm. Indeed, nanospheres in the size range between ~2 nm and ~8 nm have been synthesized and characterized in the last years[42,43,44,45,46,47] to be used for various applications.

The TiO$_2$ nanosphere models have been carved from large bulk anatase supercells and contain from 300 to almost 4000 atoms. The search for a meaningful global minimum is not feasible at the DFT level. Therefore, here the recourse to the SCC-DFTB is crucial since, as mentioned above, it allows for an accuracy that is comparable to ab initio methods, at a reduced computational cost. The initial bulk-like structures have been used as starting point for a series of simulated annealing calculations at different target temperature. Our results show that nanospheres obtained with the annealing process are generally very stable, presenting crystalline core and rearranged surfaces. The most stable structures have been then optimized at both the DFTB and hybrid DFT level of theory (even the large ones of about 4000 atoms). Such calculations are extremely demanding with hybrid DFT methods. On the basis of these results we could perform a comparative analysis of structural and electronic properties, with reference to all available experimental data. Our conclusion is, once again, also for this type of complex nanosized systems, that SCC-DFTB method is extremely efficient and more than satisfactorily accurate to obtain equilibrium structures from molecular dynamics runs, which present structural and electronic properties in very good agreement with DFT results and, when available, with experimental data.

The paper is organized as follows: in Section 2 we summarize the Computational Details, in Section 3 we present our results starting from the description of the NSs modeling and the simulated annealing process used to obtain equilibrium structures (Section 3.1). In Section 3.2 we focus on the structural analysis of the NSs, i.e. their surface distortion and undercoordinated atoms distribution. In Section 3.3, the NSs surface energies are compared. Finally, we discuss the electronic properties (in Section 3.4). All the results are summarized in the Conclusions (Section 4).

2. COMPUTATIONAL DETAILS

In this study we used two different levels of theory: SCC-DFTB and DFT. The simulated annealing processes (molecular dynamics), geometry optimizations and electronic structure calculations have been carried out using the self-consistent charge density functional tight-binding (SCC-DFTB) approach. Further geometry optimizations and electronic structure calculations have been performed also with hybrid density functional theory (DFT).

2.1 BASICS OF SCC-DFTB METHOD

The SCC-DFTB method is an approximated DFT-based method which derives from the second-order expansion of the KS-DFT total energy with respect to the electron density fluctuations. Using this assumption, the SCC-DFTB total energy can be defined as:

$$E_{tot}^{SCC-DFTB} = \sum_i \varepsilon_i + \frac{1}{2}\sum_{\alpha\beta} E_{rep,\alpha\beta}(R_{\alpha\beta}) + \frac{1}{2}\sum_{\alpha\beta} \gamma_{\alpha\beta}\Delta q_\alpha \Delta q_\beta. \qquad (1)$$

where the first term is the attractive tight-binding energy, in which the one-electron energies $\varepsilon_i$ come from the diagonalization of an approximated Hamiltonian matrix, $E_{rep,\alpha\beta}(R_{\alpha\beta})$ is a pairwise distance-dependent repulsive potential for the pair of atoms α and β, which approximates the short-range repulsion term, $\Delta q_\alpha$ and $\Delta q_\beta$ are the induced charges on the atoms α and β, respectively, and $\gamma_{\alpha\beta}$ is a Coulombic-like interaction potential.

Further information about the SCC-DFTB method can be found in the Section S1 of the supplementary material and in Refs. 24, 48 and 49. From now on, DFTB will be used as shorthand for SCC-DFTB.

2.2 ELECTRONIC STRUCTURE CALCULATIONS

For all the DFTB calculations, the open-source simulation package DFTB+ has been used.[50] We made used of the well-suited "matsci-0-3" set of parameters as reported in Ref. 35. For conjugate-gradient relaxations and electronic structure evaluations, the convergence threshold on the self-consistent charge (SCC) procedure was kept to $10^{-6}$ au and forces were relaxed to less than $10^{-4}$ au. Born-Oppenheimer DFTB molecular dynamics, used to simulate the temperature-annealing processes, was performed within the canonical ensemble (*NVT*). The Newton's equations of motion were integrated with the Velocity Verlet algorithm and a relative small time step of 0.5 fs ensured reversibility. A Nósé-Hoover thermostat, with time constant of 0.03 ps, was use to target the desired temperature during the simulated annealing simulations.

All the DFT geometry optimizations and electronic structure calculations were performed with the massive parallel version of the CRYSTAL14 code,[51] where the Kohn−Sham orbitals are expanded in Gaussian-type orbitals (the all-electron basis sets are O 8-411(d1), Ti 86-411(d41) and H 511(p1)). All forces were relaxed to less than $4.5 \cdot 10^{-4}$ au, except for the largest nanosphere (3873 atoms) for which the mean forces value was relaxed to less than $1.7 \cdot 10^{-3}$ au, due to the high computational cost. The B3LYP hybrid functional[52,53] has been used throughout this work in order to correctly describe the electronic structure of the anatase $TiO_2$.

For bulk anatase calculations with the DFTB method, we used a $16 \times 16 \times 16$ Monkhorst–Pack grid for k-point sampling. The optimal lattice parameters of the unit cell were obtained using the lattice optimization algorithm, as implemented in the DFTB+ code. Bulk anatase lattice parameters, as obtained with B3LYP functional, have been taken from a previous work by some of us (see Table I).[54] Calculated DFTB values are in very good agreement with the DFT and experimental references.[55]

**Table I.** Anatase $TiO_2$ bulk $a, c$ lattice parameters (in Å), lattice parameters ratio $c/a$, axial ($d_{ax}$) and equatorial ($d_{eq}$) Ti-O distances (in Å). DFTB values are given, together with experimental data, standard GGA-DFT(PBE) and hybrid-functional DFT(B3LYP) ones.

| Method | $a$ | $c$ | $c/a$ | $d_{ax}$ | $d_{eq}$ |
|---|---|---|---|---|---|
| DFTB This work | 3.810 | 9.732 | 2.554 | 1.995 | 1.955 |
| Exp. [55] | 3.782 | 9.502 | 2.512 | 1.979 | 1.932 |
| PBE [56] | 3.789 | 9.612 | 2.537 | 1.995 | 1.938 |
| B3LYP [54] | 3.789 | 9.777 | 2.580 | 2.000 | 1.946 |

To describe the anatase (101) surface, we used a $2 \times 2$ supercell model consisting of ten triatomic layers and 120 atoms. Periodicity was considered only along the $[10\bar{1}]$ and $[010]$ directions, while no periodic boundary conditions were imposed in the direction perpendicular to the surface. A Monkhorst-Pack k-point mesh of $2 \times 2 \times 1$ ensured the convergence of the electronic structure for both DFTB and DFT(B3LYP) calculations.

For both DFTB and DFT(B3LYP) calculations, simulated total densities of states (DOS) of the nanoparticles have been obtained through the convolution of Gaussian peaks ($\sigma = 0.005$ eV) centered at the Kohn-Sham energy eigenvalue of each orbital. Projected densities of states (PDOS) have been obtained by using the coefficients in the linear combination of atomic orbitals (LCAO) of each molecular orbital: summing the squares of the coefficients of all the atomic orbitals centered on a certain atom type results, after normalization, in the relative contribution of each atom type to a specific eigenstate. Then, the various projections are obtained from the convolution of Gaussian peaks with heights that are proportional to the relative contribution. The zero energy for all the DOS is set to the vacuum level, corresponding to an electron at an infinite distance from the surface.

### 2.3 STRUCTURAL ANALYSIS

Similarly to simulated PDOS, the extended X-ray adsorption fine structure (EXAFS) simulated spectra in the direct space have been constructed with the Gaussian convolution of peaks ($\sigma = 0.0005$ Å) centered at the distance lengths between each Ti atom and other atoms (O or Ti) from its first, second, and third coordination shell. Projections have been performed by taking into account only specific titanium atoms with a certain coordination sphere.

We have evaluated the Connolly surface areas ($S_{Conn}$) using the algorithm developed by Connolly.[57,58] The procedure consider a first step where the molecular surface is constructed from the overlap of all the atomic van der Waals spheres (without taking into account the few OH groups, which are not considered part of the oxide surface), then a probe sphere (with a chosen radius of 3.0 Å) is rolled on the former surface defining a series of contact points. These are used to form arcs that smooth the van der Waals surface, resulting in the Connolly surface.

We have also reported the surface-to-bulk ratio, defined as the ratio between the number of Ti and O atoms at the NS surface ($Ti_{6c\_sup}$, $Ti_{5c}$, $Ti_{4c}$, $Ti_{4c}(OH)$, $Ti_{3c}(OH)$, OH, $O_{2c}$, $O_{3c\_sup}$) and the number of Ti and O atoms in the bulk ($Ti_{6c}$ and $O_{3c}$). Undercoordinated atoms are directly deemed part of the surface. Fully or six-fold coordinated Ti ($Ti_{6c\_sup}$) are considered as superficial (sup) atoms when they are connected to at least one $O_{2c}$, while three-fold coordinated O ($O_{3c\_sup}$) are considered surface (sup) atoms when they are connected with a superficial Ti atom (i.e. $Ti_{6c\_sup}$, $Ti_{5c}$, $Ti_{4c}$, $Ti_{4c}(OH)$, $Ti_{3c}(OH)$).

## 3. RESULTS AND DISCUSSION

### 3.1 SIMULATED ANNEALING OF TiO$_2$ NANOSPHERES

The nanospheres considered in this work have been directly carved from a bulk anatase $TiO_2$ supercell. We applied different carving radii in order to obtain nanoparticles of about 1.5 nm, 2.2 nm, 3.0 nm and 4.4 nm diameter size, as described in detail in a previous work by some of us.[54] The structures have been kept stoichiometric and the too low-coordinated Ti and O atoms of the surface have been saturated with dissociated $H_2O$ molecules: three-fold and some four-fold Ti atoms were coordinated to hydroxyl groups; whereas mono-coordinated O atoms were saturated with H atoms. With this procedure we achieved what can be consider chemically stable $TiO_2$ models in line with what observed experimentally. It was reported that some water adsorbates on spherical nanoparticles cannot be completely eliminated even after annealing at 600 K.[59] This can be considered as intrinsic dissociated water that we have modeled in our NSs.

Once the models were carved and saturated as described, we have performed a conjugate-gradient geometry optimization at both DFTB and DFT level of theory to obtain a reference total energy for a geometrically constructed nanoparticle (we refer to them as "0 K" or "as carved" geometries).

In Fig. 1 we report the representation of the most stable DFT optimized nanospheres considered in this work. Color-coding refers to the different coordination of the Ti atoms and will be discussed exhaustively in Section 3.2.1.

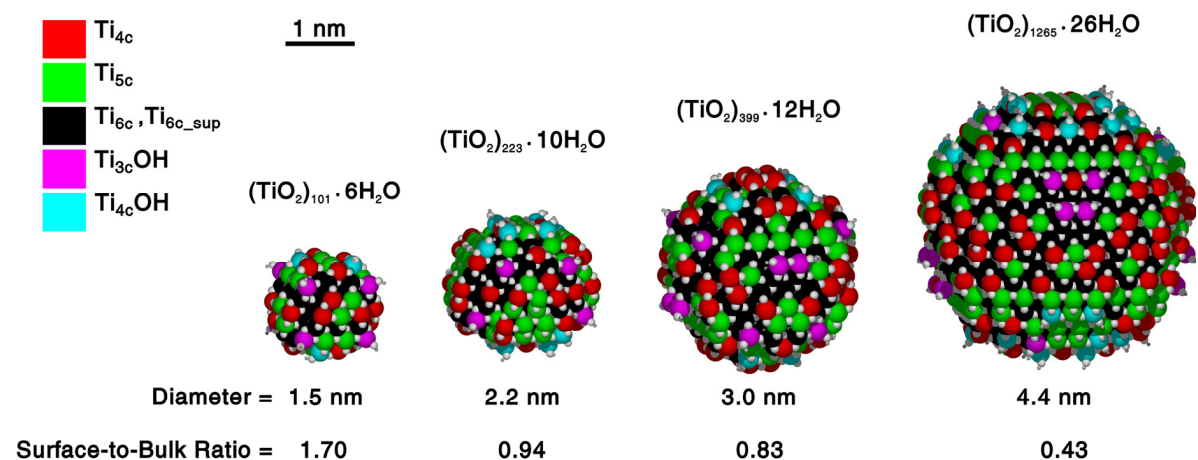

FIGURE 1. DFT(B3LYP) optimized structures, after simulated annealing, of the different nanospheres considered in this work. For each one, the stoichiometry, the approximate diameter, the surface-to-bulk ratio (as defined in the Computational details) and the position of the Ti atoms with different coordination are reported.

Successively, DFTB molecular dynamics simulations were carried out in order to simulate a temperature annealing process. We used three different target temperatures (300 K, 500 K and 700 K) for the 1.5 nm, 2.2 nm and 3 nm diameter sized nanoparticles, while the biggest nanosphere (4.4 nm) has been annealed only at 500 K. For the latter nanosphere, the cooling process has been stopped at 80 K when no further structural changes were expected. During the molecular dynamics the temperature profile has been divided in three temporal regions:

I. Heating region: the systems have been heated up to the target temperature in a very short time, starting from an initial Boltzmann velocity distribution generated at 150 K.
II. Equilibration region: the systems have been equilibrated to the target temperature until the temperature profile was found flat.
III. Cooling region: the systems have been cooled down to 0 K (80 K for the 4.4nm nanosphere) as slow as possible (depending on the size of the nanoparticles considered) in order to find the global minimum.

The representations of the simulated annealing temperature profiles for all the nanoparticles considered are reported in Fig. 2. The permanence of the system in each temporal region depends on two main factors: the temperature to be reached and the size of the NSs. The same two factors regulate the temperature gradient considered in the heating and, most of all, the cooling process: the higher is the target temperature and the bigger is the NS, the higher are the heating and cooling gradients (see Table II). Note also that, according to statistical physics, the bigger is the system, the lower are the temperature variations since thermodynamic variables fluctuate proportionally to $1/\sqrt{N}$, where $N$ is the number of atoms.

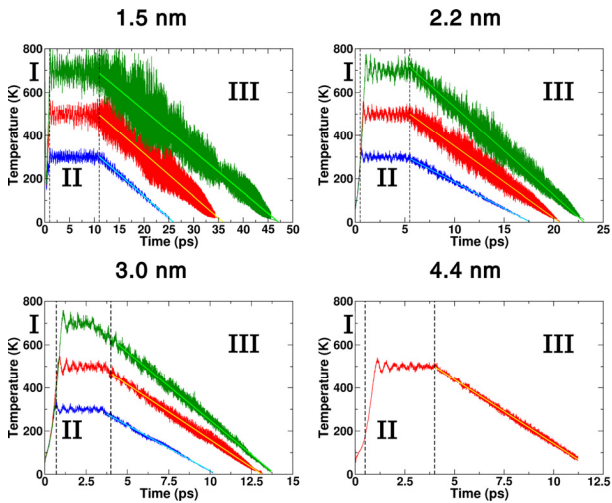

FIGURE 2. Simulated annealing temperature profiles for the 1.5 nm, 2.2 nm, 3.0 nm and 4.4 nm nanospheres. Green curves refer to the profile with 700 K of target temperature, red curves refer to the profile with 500 K of target temperature and blue curves refer to the profile with 300 K of target temperature. Heating (I), equilibration (II) and cooling (III) regions have been indicated.

**Table II.** Target temperature $T$ (in K), molecular dynamics time permanence in region I, II, II (in ps) and cooling gradient $\nabla T$ (in K/ps) of the simulated annealing processes for each nanosphere size. The target temperatures, the permanence in each region and the cooling gradients refer to the temperature profiles plots in Fig. 2.

| NS diameter | Target $T$ (K) | Permanence in Region I (ps) | Permanence in Region II (ps) | Permanence in Region III (ps) | $\nabla T$ (K/ps) |
|---|---|---|---|---|---|
| 1.5 nm | 300 | 1.0 | 10.0 | 15.0 | -20 |
|  | 500 | 1.0 | 10.0 | 25.0 | -20 |
|  | 700 | 1.0 | 10.0 | 35.0 | -20 |
| 2.2 nm | 300 | 0.25 | 5.0 | 12.30 | -24 |
|  | 500 | 0.575 | 5.0 | 15.07 | -33 |
|  | 700 | 0.925 | 5.0 | 17.28 | -40 |
| 3.0 nm | 300 | 0.25 | 3.5 | 6.41 | -41 |
|  | 500 | 0.575 | 3.5 | 9.10 | -51 |
|  | 700 | 0.925 | 3.5 | 9.27 | -63 |
| 4.4 nm | 500 | 0.5 | 3.5 | 7.2 | -59 |

Finally, one selected structure for each nanosphere obtained with the simulated annealing processes, has been optimized again both at the DFTB and DFT(B3LYP) levels of theory (the equilibrium geometries are reported in Fig. S2 in the supplementary material and Fig. 1, respectively). We selected the most stable one in terms of energy difference with respect to the not annealed "0 K" or "as carved" NSs (see Table III). The smallest 1.5 nm and 2.2 nm NSs result more stable when annealed at 300 K. At higher temperature, the surface presents some localized amorphous areas, causing a loss in stability. In the case of the 3.0 nm NS a similar surface geometry is obtained with all three simulated annealing processes at 300 K, 500 K and 700 K. We selected the 3 nm NS obtained at 500 K, which is slightly more stable than the others. For the biggest 4.4 nm NS, we decided to use a single target temperature of 500 K to reduce the computational effort. This analysis leads to the conclusion that larger NSs keep crystallinity up to higher T of annealing than smaller ones.

Values, calculated at both DFTB and DFT(B3LYP) levels of theory, of the total energy differences ($\Delta E_{DFTB}$ and $\Delta E_{DFT(B3LYP)}$) and the energy difference per TiO$_2$ formula unit ($\Delta E_{DFTB}^{unit}$ and $\Delta E_{DFT(B3LYP)}^{unit}$) between the most stable NSs obtained from the simulated annealing and the not annealed "as carved" NSs, are reported in Table III. These selected most stable NSs will be considered in the rest of this work. The DFT(B3LYP) value for 4.4 nm is not reported since we have not optimized the relative "as carved" NS, to limit the computational cost.

**Table III.** DFTB and DFT(B3LYP) relative stability (expressed in eV) of the most stable nanospheres obtained from simulated annealing processes and the respective optimized "as carved" nanospheres, as obtained from total and per TiO$_2$ formula unit energies. Since the "as carved" NS have been not optimized for the 4.4 nm, we could not compare the relative energies and thus the difference is not available (N.A.).

| NS diameter | Target $T$ | $\Delta E_{DFTB}$ | $\Delta E_{DFTB}^{unit}$ | $\Delta E_{DFT(B3LYP)}$ | $\Delta E_{DFT(B3LYP)}^{unit}$ |
|---|---|---|---|---|---|
| 1.5 nm | 300 K | -1.36 | -0.0135 | -0.17 | -0.0017 |
| 2.2 nm | 300 K | -4.76 | -0.0213 | -0.65 | -0.0029 |
| 3.0 nm | 500 K | -6.83 | -0.0171 | -0.25 | -0.0006 |
| 4.4 nm | 500 K | -10.09 | -0.0080 | N.A. | N.A. |

### 3.2 STRUCTURAL ANALYSIS OF TiO$_2$ NANOSPHERES

#### 3.2.1 SIZE AND MORPHOLOGY

The DFT(B3LYP) equilibrium geometries together with a visual representation of the Ti atoms coordination for the most stable nanospheres are reported in Fig. 1, whereas the corresponding geometries for DFTB are in Fig. S2 in the supplementary material. As one would expect, we observed that the bigger the NS, the lower the surface-to-bulk ratio (defined in the Computational Details), due to the fact that the inner bulk-like atoms increase in number.

In Table IV we report the coordination patterns for the different types of Ti atoms present in each NS, their number and the relative percentage with respect to the total number of Ti atoms. For both DFTB and DFT(B3LYP) optimized structures the agreement between the two sets of data is very good, especially for the NSs with surface-to-bulk ratio ≤ 1. Thus, we may conclude that the DFTB method is capable of catching the connectivity within the various NSs with the accuracy of DFT.

As a general comment, one may note that, as the dimension of the nanosphere increases, the amount of $Ti_{3c}(OH)$ and $Ti_{4c}(OH)$ species, required to saturate the highly undercoordinated surface Ti atoms, decrease in percentage. Superficial ($Ti_{4c}$, $Ti_{5c}$, $Ti_{6c\_sup}$) atoms grow roughly quadratically with the radius of the NS (since the NS surface area is proportional to $r^2$), whereas bulk $Ti_{6c}$ species grow roughly with the third power of the size (since the NS volume is proportional to $r^3$). Therefore, approaching the bulk limit, we have that the ratio between surface and bulk atoms tends to zero, as one would expect.

Due to the fact that the 2.2 nm NS is the smallest one presenting a surface-to-bulk ratio below the value of 1 (see Fig. 1), we will use it as the system of choice for further analysis, in the next sections.

**Table IV.** Number of Ti atoms with a specific coordination and their percentage with respect to the total number of Ti atoms for all the different size NSs geometries optimized with DFTB. Number and percentage of Ti atoms in NSs geometries optimized with DFT(B3LYP) are reported in parenthesis, when different from DFTB.

| DFTB [DFT(B3LYP)] | Number | % | Number | % | Number | % | Number | % |
|---|---|---|---|---|---|---|---|---|
| Ti site | 1.5 nm | | 2.2 nm | | 3.0 nm | | 4.4 nm | |
| $Ti_{4c}$ | 20 [19] | 19.8 [18.8] | 36 | 16.1 | 53 | 13.3 | 106 | 8.4 |
| $Ti_{5c}$ | 20 [21] | 19.8 [20.8] | 43 [49] | 19.2 [22.0] | 69 [65] | 17.3 [16.3] | 159 | 12.6 |
| $Ti_{6c\_sup}$ | 20 | 19.8 | 28 [24] | 12.6 [10.8] | 72 [75] | 18.0 [18.8] | 157 | 12.4 |
| $Ti_{6c}$ | 29 | 28.7 | 96 [94] | 43.1 [42.1] | 181 [182] | 43.4 [45.6] | 791 | 62.5 |
| $Ti_{3c}(OH)$ | 8 | 7.9 | 8 | 3.6 | 16 | 4.0 | 20 | 1.6 |
| $Ti_{4c}(OH)$ | 4 | 4.0 | 12 | 5.4 | 8 | 2.0 | 32 | 2.5 |

Another important morphological aspect is given by the variety of coordination types at the NS surface. The relative percentage of each is reported in Table V. It can be observed that, as the NS size increases, there is a general trend of reduction of the highly undercoordinated $Ti_{4c}$, $Ti_{3c}(OH)$ and $Ti_{4c}(OH)$ species in parallel to an increase of the number of $Ti_{5c}$ and $Ti_{6c\_sup}$ atoms.

**Table V.** Percentage of under-coordinated Ti atoms composing the surface of all the different size NSs geometries optimized with DFTB. Percentage of under-coordinated Ti atoms composing the surface of NSs geometries optimized with DFT(B3LYP) are reported in parenthesis, when different from DFTB.

| DFTB [DFT(B3LYP)] | Surface Composition (%) | | | |
|---|---|---|---|---|
| Ti site | 1.5 nm | 2.2 nm | 3.0 nm | 4.4 nm |
| $Ti_{4c}$ | 27.8% [26.4%] | 28.3% [27.9%] | 24.3% [24.4%] | 22.5% |
| $Ti_{5c}$ | 27.8% [29.2%] | 33.9% [38.0%] | 31.6% [29.9%] | 33.5% |
| $Ti_{6c\_sup}$ | 27.8% | 22.0% [18.6%] | 33.0% [34.6%] | 33.1% |
| $Ti_{3c}(OH)$ | 11.1% | 6.3% [6.2%] | 7.3% [7.4%] | 4.2% |
| $Ti_{4c}(OH)$ | 5.5% | 9.5% [9.3%] | 3.7% | 6.7% |

### 3.2.2 STRUCTURAL DISTORTIONS

Similarly to nanotubes and nanorods, the high curvature of the small nanospheres surface is the key aspect for their enhanced affinity to chemical adsorbates. In order to determine how nanostructuring influences not only the coordination type of the surface atoms, but also their geometrical environment, we systematically analyze bond length modifications with respect to bulk values.

To this end, we have simulated direct space EXAFS (X-ray absorption fine structure spectra) calculating the density of distances for each Ti atoms with the other (Ti or O) atoms and projecting them on Ti atoms with different coordination patterns. In Figure 3, results for the relaxed (after temperature annealing) DFTB and DFT(B3LYP) nanospheres with a 2.2 nm diameter size are reported. In both cases, we used four different starting geometries for full atomic relaxation: the one defined "0 K" (or "as carved") and the three configurations obtained after the simulated annealing processes at 300 K, 500 K and 700 K of target temperature. The analogous spectra for the 1.5 nm, 3.0 nm and 4.4 nm NSs are given in the supplementary material (see Fig. S3, Fig. S4 and Fig. S5, respectively).

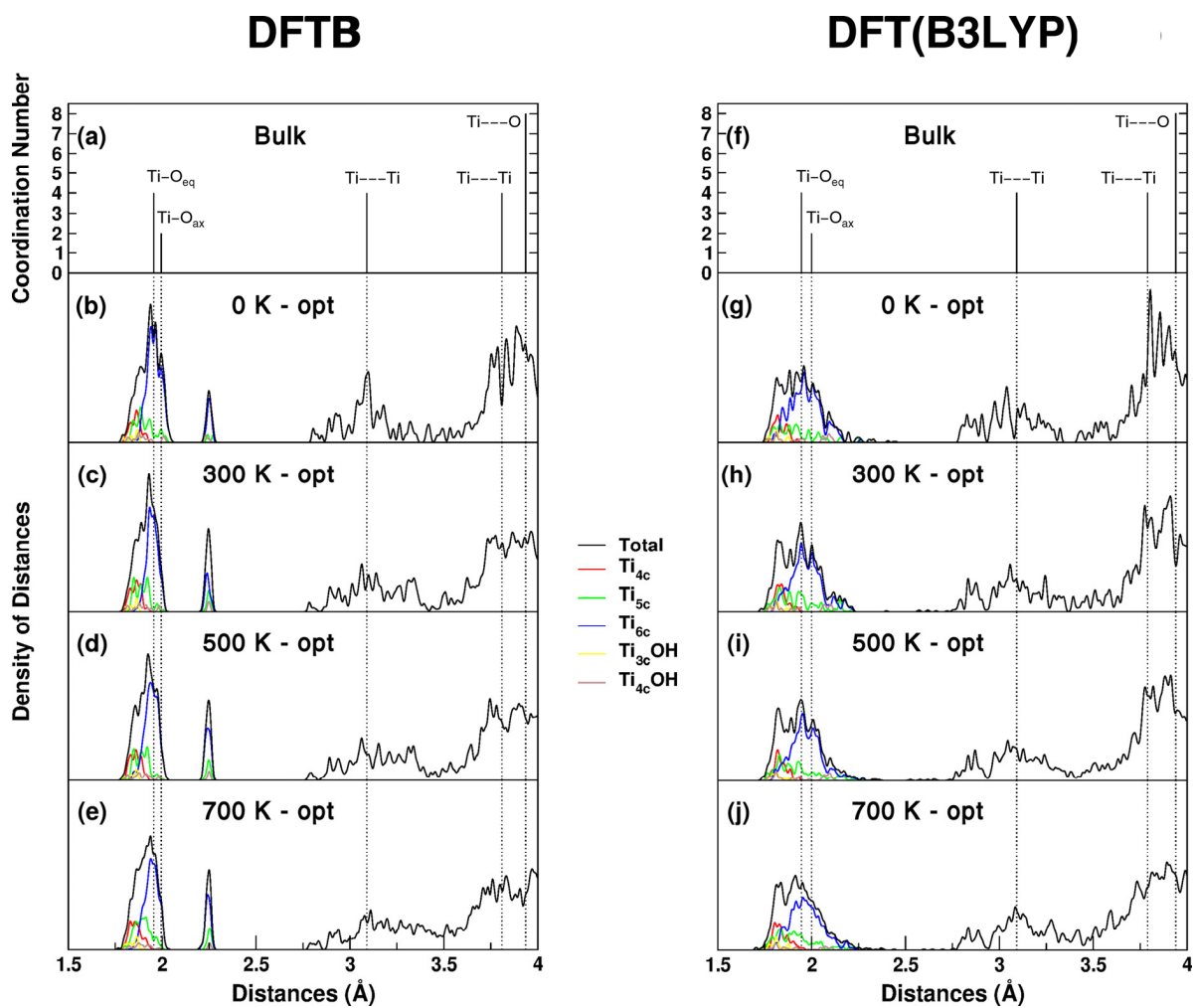

FIGURE 3. Distances distribution (simulated EXAFS) computed with DFTB and DFT(B3LYP) for bulk anatase (top panel (a), (f)), computed with DFTB for 2.2 nm NSs produced at different

temperature 0 K (b), 300 K (c), 500 K (d), 700 K (e) and computed with DFT(B3LYP) at 0 K (g), 300 K (h), 500 K (i), 700 K (j).

The top panels (Fig. 3(a) and 3(f)) report the reference lines for bulk anatase $TiO_2$ as obtained by performing structural relaxation with DFTB and DFT(B3LYP), respectively. The lines shown are assigned to different coordination spheres of a selected Ti atom: the first two lines represent the first coordination sphere, thus the Ti−$O_{eq}$ and Ti−$O_{ax}$ bonds, which are well-known to be slightly different in a $D_{2d}$ point symmetry (DFTB: Ti−$O_{eq}$ = 1.955 Å and Ti−$O_{ax}$ = 1.995 Å, DFT(B3LYP): Ti−$O_{eq}$ = 1.946 Å and Ti−$O_{ax}$ = 2.000 Å); then the third line (second coordination sphere) is the distance between the selected central Ti atom and the next-neighboring Ti atoms (Ti⋯Ti) of 3.090 Å in the DFTB structure and 3.092 Å in the DFT(B3LYP) one; finally, the fourth and the fifth lines (third coordination sphere) are the distances between the central Ti atom and the second shell of Ti and O atoms (Ti⋯Ti and Ti⋯O) of 3.809 and 3.936 for DFTB and of 3.789 Å and 3.939 Å for DFT(B3LYP), respectively.

Figs. 3(b)-3(e) report the DFTB spectra for optimized particles from increasing production temperature simulations: from the "0 K" or "as carved" NS to the 300 K, 500 K and 700 K ones. With respect to the bulk lines, we can notice a broader variety of features contributing to the peak of the first coordination sphere in all the cases. Generally, there is a predominant contribution of the $Ti_{6c}$ (blue line), whereas under-coordinated Ti atoms present shorter Ti-O distances. Regarding the second and third coordination spheres, we observe peaks centered at the bulk Ti⋯Ti and Ti⋯O distances, with some broadening due to surface species. In line with what already discussed in Section 3.1, the effect of the target temperature used for the simulated annealing is evident: as the temperature increases, the distribution of distances peaked around the Ti−$O_{eq}$ and Ti−$O_{ax}$ values in the first coordination sphere broadens, the contribution of the $Ti_{6c}$ (blue line) decreases and the one from under-coordinated Ti atoms slightly increases. This turns out into a gradual increasing of the population of Ti-O distances around 1.75 Å. A similar broadening effect can also be noticed in the second and third coordination spheres distribution of distances. Thus, the higher is the production temperature, the less crystalline is the NSs surface and the more widespread is the distances distribution.

Figs. 3(g)-3(j) report the corresponding DFT(B3LYP) spectra for the same particles considered above. The qualitative picture is essentially the same as for DFTB, with a predominant contribution of the $Ti_{6c}$, with shorter distances for under-coordinated Ti atoms and with a decreasing crystallinity as the production temperature increases. However, in this case, we have a finer description of the distances, in particular there is a broad and continuous distribution of the Ti-O distances, in the first coordination sphere, that are larger than the Ti−O$_{ax}$ one. In the DFTB case they tend to concentrate at the same value of about 2.25 Å. In other words, the DFTB optimized NSs correctly contain some Ti-O bonds that are longer than 2.05 Å, but, unlike DFT, their equilibrium distance tends to be always about the same (~2.25 Å).

One should be careful before attributing the failure to the DFTB approximation because DFTB has been parameterized to PBE calculations and not to B3LYP ones. To clarify this point, we have performed a DFT(PBE) optimization of the nanoparticle and compared the DFT(PBE) EXAFS spectrum with those by DFTB and DFT(B3LYP) (see Fig. S6 in supplementary material). Indeed DFT(PBE) and DFT(B3LYP) are quite similar, whereas DFTB presents the different behavior described above in the range between 2.0-2.5 Å. This discrepancy is related to the tight-binding composition of the total energy made up by a repulsive and an attractive term (see Computational Details): on the one hand, the two-center repulsive potential is exactly the same for all Ti-O pairs (see Fig. S1 in the supplementary material), steeply descending until 2.25 Å, where it is already zero; on the other hand, the attractive term, is steadily decreasing and in the range between 2.05 and 2.25 Å is not large enough to compensate for the repulsive term. Therefore, Ti-O bonds longer than 2.25 Å are not favored because of a smaller attractive term and a flat repulsive one, whereas Ti-O bonds shorter than 2.25 Å are not favored because the attractive term is smaller than the repulsive one (at least in the range between 2.05 and 2.25 Å). Of course, this is a limitation in the description of the fine structural details of the nanoparticles by DFTB method, however, it is not that severe and it will only have a little effect on the electronic structure, as discussed below in Section 3.4.2.

### 3.3 SURFACE ENERGY

In this section we evaluate the cost to form a nanosphere from the bulk system in terms of surface energy, in order to understand the energetics related to the NSs curvature. Considerations on the standard free energy of formation of a nanosphere have been already tackled in a previous work of some of us.[54] Therefore, as commonly done, we will approximate the total free energy of formation ($G_{NS}^{surface}$) using the total energy ($E_{NS}^{surface}$). We evaluated the surface area with the Connolly procedure ($S_{Conn}$) for each NS and we refer to the surface energy as $\gamma = E_{NS}^{surface}/S_{Conn}$. In Table VI we compare the $S_{Conn}$ and surface energy $\gamma$ for each NS calculated at DFTB and DFT(B3LYP) level of theory with the one of the regular (101) anatase surface.

One may notice that, on one side, the Connolly surfaces ($S_{Conn}$) are very similar for NSs optimized with the two methods, with the DFT ones slightly lower than the DFTB. On the other, DFTB tends to overestimate surface energies ($E_{NS}^{surface}$), thus resulting γ values are overestimated (by ~0.5 J/m$^2$) with respect to the DFT ones. However, it is also evident that with both methods the surface energy does not dependent on the NS size. This is due to the compensation of two factors. On the one hand, the bigger is the nanoparticle, the lower is the number of highly undercoordinated superficial atoms and so the surface energy should decrease. However, on the other hand, since the additional water molecules we used to saturate the extremely low coordinated sites ($Ti_{3c}$ and $O_{1c}$), or intrinsic water molecules, are relatively more abundant in smaller NSs (see the percentage of $H_2O$ molecules per $TiO_2$ units, $nH_2O/nTiO_2$, in Table VI), there is an energy compensation effect, which results in about the same surface energy. A previous study, based on the less accurate molecular mechanics approach and not considering the intrinsic water, failed in the correct description of the surface energy trend that was found to increase with the NSs size.[60]

**Table VI.** Connolly surface area ($S_{Conn}$ in nm$^2$) and Surface energy (γ in J/m$^2$) calculated for different size NSs optimized with DFTB and DFT(B3LYP).

| NS | 1.5 nm $(TiO_2)_{101} \cdot 6\ H_2O$ | | 2.2 nm $(TiO_2)_{223} \cdot 10\ H_2O$ | | 3.0 nm $(TiO_2)_{399} \cdot 12\ H_2O$ | | 4.4 nm $(TiO_2)_{1265} \cdot 26\ H_2O$ | | (101) Surface |
|---|---|---|---|---|---|---|---|---|---|
| $nH_2O/nTiO_2$ | 5.9% | | 4.4% | | 3.0% | | 2.0% | | |
| Method | $S_{Conn}$ | γ | $S_{Conn}$ | γ | $S_{Conn}$ | γ | $S_{Conn}$ | γ | γ |
| DFTB | 13.03 | 1.23 | 22.42 | 1.25 | 33.13 | 1.23 | 70.29 | 1.27 | 0.71 |
| DFT(B3LYP) | 12.76 | 0.70 | 22.11 | 0.72 | 32.54 | 0.70 | 69.48 | 0.74* | 0.54 |

*Partially optimized

### 3.4 ELECTRONIC STRUCTURE

#### 3.4.1 BULK PROPERTIES

In this and in the next section we aim, on one side, to evaluate how accurate is the DFTB method with respect to a hybrid DFT, such as B3LYP, in describing the electronic properties of bulk and nanostructured TiO$_2$ and, on the other, to understand how nanostructuring affects those properties.

Since TiO$_2$ is a fundamental material in photo(electro)chemistry, photocatalysis, photovoltaics, many reliable theoretical and experimental data are already available in the literature, regarding the electronic structure of this indirect gap semiconductor. In this part of our work we will compare results obtained at DFTB level of theory with those from the more accurate, but also costly, DFT-based techniques. In Table VII together with our calculated DFTB band gap of bulk anatase, we reported PBE,[61] PBE0,[62] B3LYP,[54] PBE+U,[62] GW[63] and experimental data measured at temperature close to 0 K.[64] The agreement between DFTB, experimental data and Hubbard corrected values is extremely good, probably for error cancellation.[34] As widely known, DFT-GGA methods severely underestimate the anatase TiO$_2$ band gap, while the introduction of some exact exchange in hybrid functionals (i.e. PBE0, B3LYP, HSE06) provides more accurate values. However, it is not yet clear whether this discrepancy is due to excitonic effects not included in the calculations, since even more sophisticated GW calculations provide slightly overestimated band gaps close to those obtained with hybrid functionals.[63]

**Table VII.** Electronic band gap (expressed in eV) for bulk anatase TiO$_2$. DFTB values have been compared to experimental data, values calculated with standard GGA-DFT, with hybrid DFT functionals, GGA-DFT Hubbard corrected and GW.

| Method | Band Gap |
|---|---|
| Exp. [64] | 3.4 |
| DFTB This work | 3.22 |
| PBE [61] | 2.36 |
| PBE0 [62] | 4.50 |
| B3LYP [54] | 3.81 |
| PBE+U [62] | 3.27 |
| GW [63] | 3.83 |

3.4.2 NANOPARTICLES DENSITY OF STATES

The analysis of nanoparticles electronic structure is more complex. These are molecular systems of finite size, therefore, one cannot define true band states and band gaps. However, it is possible to distinguish between "very localized" states (i.e. molecular orbitals, MOs) and states that are delocalized on several atoms of the nanoparticle, similarly to periodic systems (i.e. "pseudo band" states). Thus, two different kinds of energy gaps can be defined: the canonical HOMO-LUMO gap ($\Delta E_{H-L}$), which is the energy difference of the frontier orbitals, and the Kohn-Sham band gap ($\Delta E_{gKS}$), which is the energy difference of the highest occupied delocalized valence state and the lowest unoccupied delocalized conduction state.

The definition of delocalized or "pseudo band" state is not straightforward. In general, a localized state is expected to present one or few atomic orbital coefficients with a relatively high value and all the others being almost zero, whereas a delocalized state is expected to have several atomic orbital coefficients with comparable values. For the nanospheres under study, frontier orbitals and few next eigenstates are quite localized at the surface. Delocalized states are spatially located in the inner part of the nanospheres, with deeper energies in the valence and higher ones in the conduction region, respectively. Nonetheless, we do not observe a net transition from localized to delocalized states. Thus, we must define a criterion based on the maximum squared coefficient ($max_c$) of each eigenstate: the first delocalized state is the one (in the valence and conduction region, respectively) with $max_c$ lower than a threshold of 0.02. In this way, $\Delta E_{gKS}$ is rigorously defined and values for different NSs and for different methods can be compared.

**Table VIII.** HOMO-LUMO electronic gap ($\Delta E_{H-L}$) and Kohn-Sham electronic gap $\Delta E_{gKS}$ (expressed in eV) calculated for different size NSs with both DFTB and DFT(B3LYP) methods and with DFTB on top of the DFT(B3LYP) optimized geometries (DFTB/DFT(B3LYP)).

| NS Diam. | $\Delta E_{H-L}$ | | | $\Delta E_{gKS}$ | | |
|---|---|---|---|---|---|---|
| | DFTB | DFT (B3LYP) | DFTB/ DFT(B3LYP) | DFTB | DFT (B3LYP) | DFTB/ DFT(B3LYP) |
| 1.5 nm | 3.12 | 4.23 | 3.43 | 3.62 | 4.81 | 4.21 |
| 2.2 nm | 3.11 | 4.13 | 3.38 | 3.55 | 4.31 | 3.84 |
| 3.0 nm | 2.95 | 4.00 | 3.27 | 3.42 | 4.13 | 3.45 |
| 4.4 nm | 2.95 | 3.92 | 3.23 | 3.33 | 3.96 | 3.36 |

In Table VIII we report the calculated values of $\Delta E_{H-L}$ and $\Delta E_{gKS}$ with DFT(B3LYP) and with DFTB ones. Furthermore, since DFTB method was found to fail in the fine description of some Ti-O distances, as discussed in Section 3.2.2, we also present single-point DFTB electronic structure calculations on the DFT optimized structures. We refer to these results as DFTB/DFT(B3LYP) and we use them to test how DFTB behaves in the description of the electronic structure when a more accurate geometry is used.

Starting our analysis from the $\Delta E_{H-L}$ gaps in Table VIII, we can see that all the methods show a gradual reduction of the HOMO-LUMO gap with size. However, DFT(B3LYP) values are always about ~1.0/1.2 eV higher with respect to the DFTB ones, but only ~0.7/0.8 eV higher if we compare with the DFTB/DFT(B3LYP) results. This discrepancy is in part due to the intrinsic difference in the description of the bulk anatase TiO$_2$ band gap between the two methods (see Table VII), of about 0.6 eV [DFT(B3LYP) 3.81 eV – DFTB 3.22 eV]. However, another crucial factor is the equilibrium geometry, which affects the position of the frontier MOs associated eigenvalue in the Kohn-Sham band gap or $\Delta E_{gKS}$. In the DFTB case, those states are slightly shifted from the edges of the valence band or the conduction band into the gap, leading to a smaller $\Delta E_{H-L}$. This is a consequence of the over-localization of these molecular states due to a non-accurate DFTB structural description of some Ti-O$_{ax}$ distances at the surface of the nanosphere, which are too long (about 2.25 Å, as we already discussed in Section 3.2.2).

To prove this, we must compare the DFTB band gaps values when the DFT(B3LYP) geometry is used, where, all the superficial Ti-O$_{ax}$ distances are correctly described. The localized eigenstates (MOs) are now closer to the delocalized ones that we consider to be the top of the valence and the bottom of the conduction bands. This leads to an increase in the calculated $\Delta E_{H-L}$ gap, which is now only ~0.7/0.8 eV lower than the DFT(B3LYP) one (essentially because of an intrinsic difference between the two methods that was also observed in the bulk case above).

As the NSs size increases, there is a smooth decrease of the Kohn-Sham band gap, approaching the bulk value. The difference between DFTB and DFT(B3LYP) $\Delta E_{gKS}$ values is in the range between 0.6-1.2 eV and it is reduced to ~0.5/0.7 eV when DFTB/DFT(B3LYP) and DFT(B3LYP) results are compared.

Different experimental works based on UV-Vis optical techniques or optical waveguide spectra (OWS), as well as solid-state theory, agree with the fact that the band gap of TiO$_2$ nanoparticles increases with decreasing particle size, due to quantum confinement effects.[42,65,66] In particular, Reddy and coworkers reported a band gap enhancement of 0.1-0.2 eV for a sample of TiO₂ anatase nanoparticles with a mean size range of 5-10 nm with respect to bulk.[67] Liu and collaborators measured, by means of UV-Vis spectroscopy, a blue shift of approximately 0.15 eV relative to bulk for very small (average diameter 3.0 nm) anatase TiO₂ colloidal solutions.[42] This is in line with what DFT(B3LYP) and DFTB calculations predict. We compute a quite large HOMO-LUMO gap value for the 1.5 nm NSs and we observe a trend of reduction of $\Delta E_{H-L}$ towards the bulk value, as the size of the particle increases.

In Figures 4(a)-4(b)-4(c), DFTB, DFT(B3LYP) and DFTB/DFT(B3LYP) total (DOS) and projected densities of states (PDOS) are reported for the NSs with a 2.2 nm diameter. In the lower panels of Fig. 4, the maximum squared coefficients ($max_c$) are reported for each eigenstate in the DOS. From the PDOS analysis it is evident that with all the theoretical approaches, the states at the top of the valence band, near the HOMO level, are composed essentially from "2p" orbitals of the $O_{2c}$ atoms or OH groups, i.e. the HOMO is located on the surface (red and green curves in the upper panels). At lower energies, the contribution of the $O_{3c}$ (core of the NS, blue curves in Figure 4) becomes predominant. We can note that the DFTB eigenstates near the valence band edge are much more localized ($max_c$ between ~0.25 and ~0.1) with respect to the DFT(B3LYP) ($max_c$ between ~0.1 and ~0.02) and also to the DFTB/DFT(B3LYP) ones ($max_c$ between ~0.2 and ~0.1). However, the DFTB and DFTB/DFT(B3LYP) description of the valence band is extremely satisfactory and well reproduces the DFT(B3LYP) reference calculations.

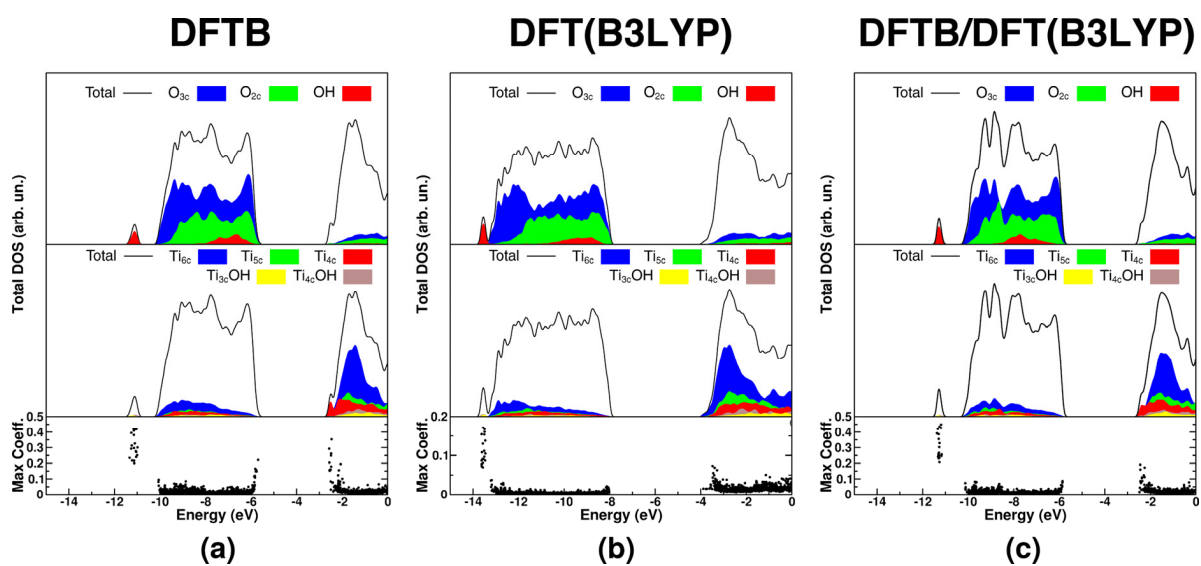

(a)  (b)  (c)

FIGURE 4. (a) DFTB, (b) DFT(B3LYP) and (c) DFTB/DFT(B3LYP) total (DOS) and projected (PDOS) density of states on different coordinated O (upper panel) and Ti (middle panel) atoms of the nanosphere. In the lower panel, the maximum atomic orbital squared coefficient ($max_c$) of each eigenstate is reported. High values of $max_c$ correspond to localized states while low values correspond to delocalized states.

The states close to the bottom of the conduction band are quite localized, as indicated by the high $max_c$ values for all the three theoretical approaches (Figure 4(a)-4(b)-4(c)). There is one exception, which is the DFT(B3LYP) LUMO level that is delocalized in the core of the NS, as indicated by the large $Ti_{6c}$ component in the PDOS (blue curves in Fig. 4(b), middle panel). In this particular case, the localized states are at higher energy, where the contribution of the $Ti_{4c}$ and $Ti_{5c}$ states (red and green curves) to the DOS became significant. With DFTB, the LUMO level and the states close to it are all localized at the surface ($Ti_{4c}$ and $Ti_{5c}$ atoms) and only at higher energy we start to observe some delocalization in the core of the NS ($Ti_{6c}$ atoms). The DFTB/DFT(B3LYP) results are intermediate between DFTB and DFT(B3LYP), which is an indication of the role played by the structure in the localization/delocalization balance.

In Figure 5 we report the total DOS for all the NSs of different size with the three theoretical approaches. From the analysis of $max_c$, we can confirm that all the NSs have localized states at the edges of the valence bands and there is a general trend for the DFTB method to over-localize them with respect to the DFTB/DFT(B3LYP) and DFT(B3LYP). Furthermore, it is worth mentioning that as the NS size increases and the electronic structure (DOS) starts to approach the one of the bulk $TiO_2$, the DFTB description is more similar to the DFT(B3LYP) one. In addition, from the comparison between DFTB and DFTB/DFT(B3LYP) results, we have further confirmation to what discussed in Section 3.2.1 and 3.2.2 regarding the poor description by DFTB of the superficial Ti-$O_{ax}$ distances peaked at 2.25 Å.

From the shape of the DOS and the $max_c$ values reported in Figure 5, one can find further support (besides EXAFS analysis) that the DFTB structural description of NSs improves as the dimension of the nanospheres increases. The larger are the NSs, the less curved is the surface, the more accurate the description. Indeed, we can observe that i) while the shape of the DOS considerably changes passing from the DFTB (Fig. 5(a)) to the DFTB/DFT(B3LYP) (Fig. 5(c)) for the small NSs (1.5 and 2.2 nm), this is not true for larger ones (3.0 and 4.4 nm); ii) the localization of the frontier eigenstates (given by $max_c$ values) is lower with DFTB (Fig. 5(a)) than DFTB/DFT(B3LYP) (Fig. 5(c)) for the small NSs, whereas it is almost the same for the larger ones.

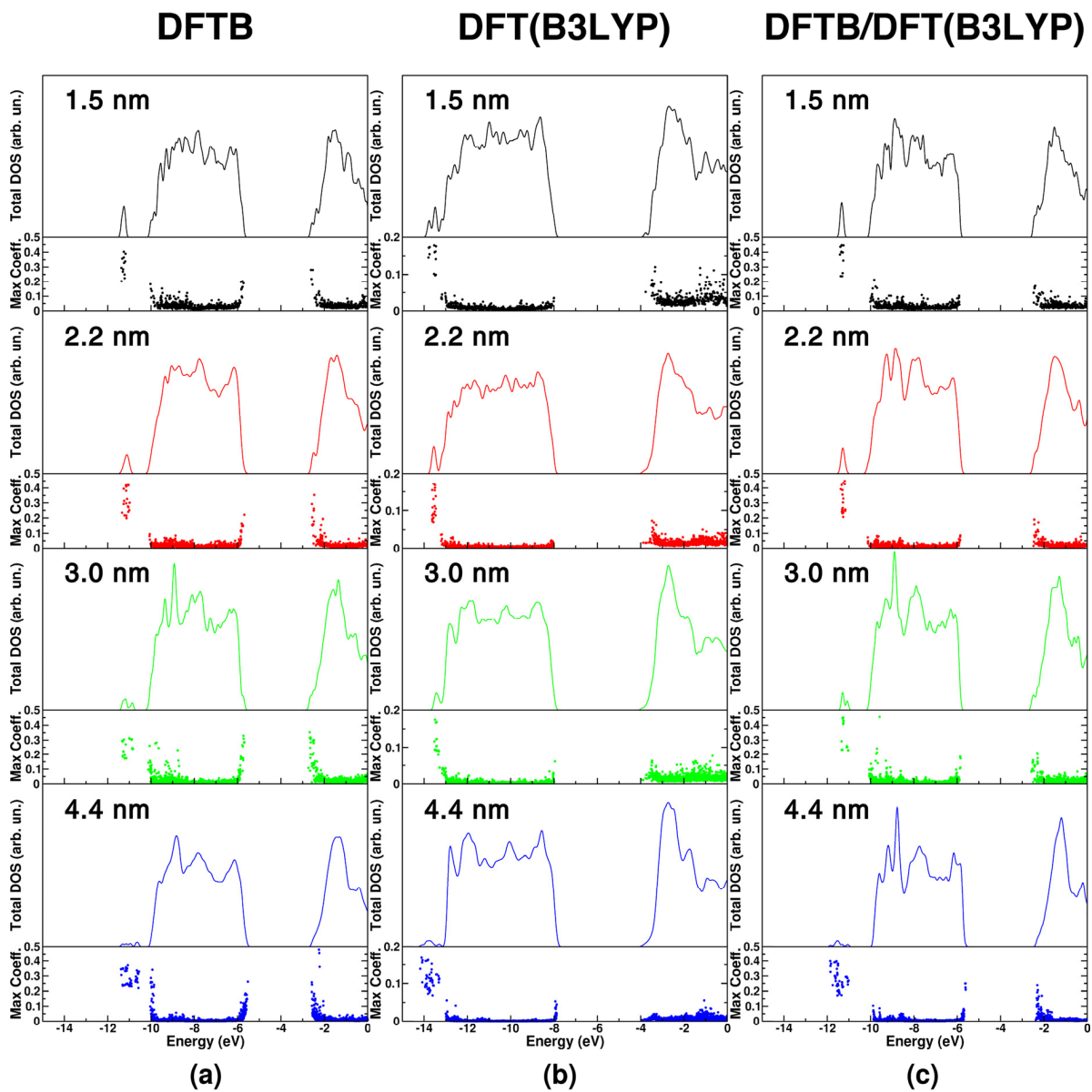

**FIGURE 5.** (a) DFTB, (b) DFT(B3LYP) and (c) DFTB/DFT(B3LYP) total (DOS) density of states for different size nanosphere, 1.5 nm (black), 2.2 nm (red), 3.0 nm (green), 4.4 nm (blue). For each NSs the DOS has been normalized to the number of $TiO_2$ units, in order to have comparable DOS intensities. Below each DOS, the maximum atomic orbital coefficient ($max_c$) of each eigenstate, as a function of the energy, is also graphically reported. High values of $max_c$ correspond to highly localized states while low values correspond to delocalized states.

The main conclusion from this section is that DFTB is quite successful in describing the electronic properties of $TiO_2$ based systems. The most critical aspect is found to be related to its description of the structural details at the surface of the nanoparticles, especially when these are very small. The elongated superficial Ti-$O_{ax}$ distances of the DFTB equilibrium structures (in particular the small NSs) bring to an over-localization of molecular states at the surface of the nanospheres. However, when the DFTB electronic structure calculations are performed on DFT(B3LYP) optimized geometries or if we consider nanosphere with a certain size, approximately with a surface-to-bulk ratio ≤ 1, which is also the most interesting experimental size range, we observe an improved picture both from a qualitative and quantitative point of view.

## 4. CONCLUSIONS

In this work we have performed a systematic study of $TiO_2$ nanospheres of increasing diameter (from 1.5 to 4.4 nm) comparing the results obtained with SCC-DFTB with those obtained with DFT(B3LYP). For the largest nanosphere we are dealing with about 4000 atoms, which is extremely demanding for first-principles calculations and achievable only by using a massive parallel version of the CRYSTAL14 code.

First of all, SCC-DFTB method has been crucial to perform some molecular dynamics simulations searching for the global minimum for these very complex and multi-configurational systems. The temperature annealing processes lead to structures which are more stable than the original "as carved" ones. The energy gain in stabilization observed with SCC-DFTB is confirmed by subsequent DFT(B3LYP) optimizations.

Surface energies of these models are not dependent on the NSs size, both with SCC-DFTB and DFT(B3LYP) methods, which has relevant experimental implications: the increased cost to make smaller nanosphere is nicely balanced by the increased relative amount of intrinsic dissociated water on the surface.

From the structural point of view, SCC-DFTB can be considered extremely successful, since it very well reproduces the experimental and DFT lattice parameters and the bulk interatomic distances (Ti-O). When going to the nanospheres we can still confirm the accurate description by SCC-DFTB of the core part, with some little discrepancies for the atomic distances at the surface. DFT(B3LYP) can better describe small variations in distances for different Ti-O whereas SCC-DFTB tends to aggregate them into a smaller range of values. The larger is the NS, the smaller the relative number of surface atoms with respect to core ones. Therefore, SCC-DFTB is progressively more accurate at increasing size. This is very good news since the computational community is interested in using such cheap computational method to describe systems of realistic size (diameter > 4 nm) that are those more commonly used in experiments.

Regarding the electronic structure, as previously observed, the bulk band gap of $TiO_2$ is extremely well described by SCC-DFTB, although probably as a consequence of some cancellation of errors. In the case of the NSs, SCC-DFTB reproduces the quantum size effect leading to a higher band gap, in line with experimental observations. Also, the trend with NS increasing size is similar to that computed with DFT(B3LYP). The SCC-DFTB DOS are very similar to the DFT(B3LYP) once, except some difference in the degree of localization of the surface states at the valence (top) and conduction (bottom) band edges.

Concluding, this study has proven that SCC-DFTB is a valuable tool for the global optimization of large anatase $TiO_2$ nanoparticles and that it provides geometries and electronic structures with quite satisfactory accuracy when compared with the corresponding DFT(B3LYP) data, especially for the case of NS with a 4.4 nm diameter size. This type of benchmark has required investing large computational resources since we had to optimize systems with up to ~4000 atoms at the hybrid DFT level of theory.

**Supplementary Material**

See supplementary material for the details on the SCC-DFTB method, plot of the DFTB repulsive term as a function of the interatomic distance, equilibrium geometry and Ti atoms coordination of the SCC-DFTB optimized NSs, distances distribution (simulated EXAFS) computed with DFTB and DFT(B3LYP) for the 1.5 nm, 3.0 nm and 4.4 nm NSs, distance distribution (simulated EXAFS) comparison between DFT(PBE), DFT(B3LYP), DFTB for the 2.2 nm NS.

**Acknowledgments**

The authors are grateful to Prof. Gotthard Seifert for very fruitful discussions and to Lorenzo Ferraro for his technical help. The project has received funding from the European Research Council (ERC) under the European Union's HORIZON2020 research and innovation programme (ERC Grant Agreement No [647020]) and from CINECA supercomputing center through the computing LI05p_GRV4CUPT grant.

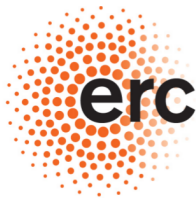
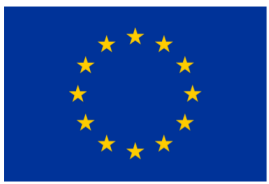

**References**
___________________

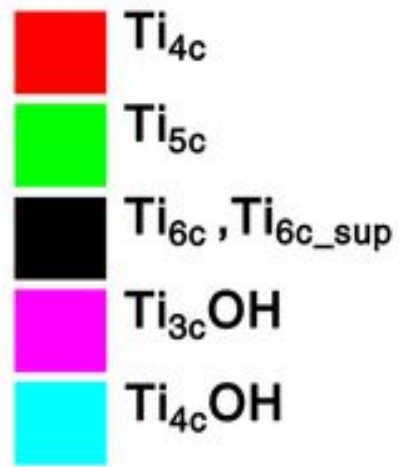
- Ti$_{4c}$
- Ti$_{5c}$
- Ti$_{6c}$, Ti$_{6c\_sup}$
- Ti$_{3c}$OH
- Ti$_{4c}$OH

1 nm

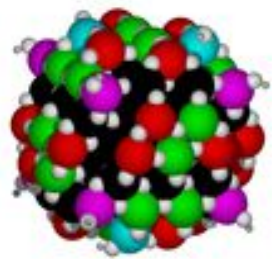

(TiO$_2$)$_{101}$ · 6H$_2$O

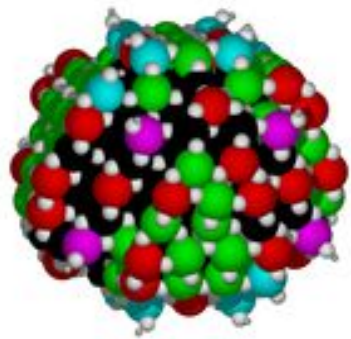

(TiO$_2$)$_{223}$ · 10H$_2$O

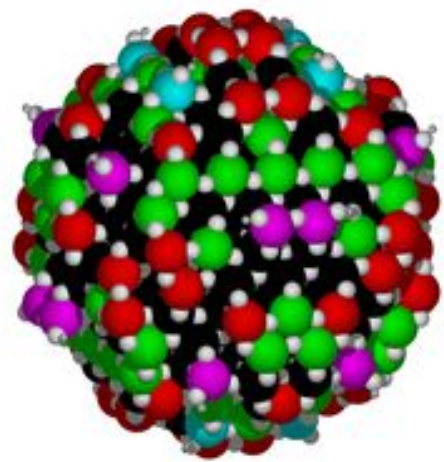

(TiO$_2$)$_{399}$ · 12H$_2$O

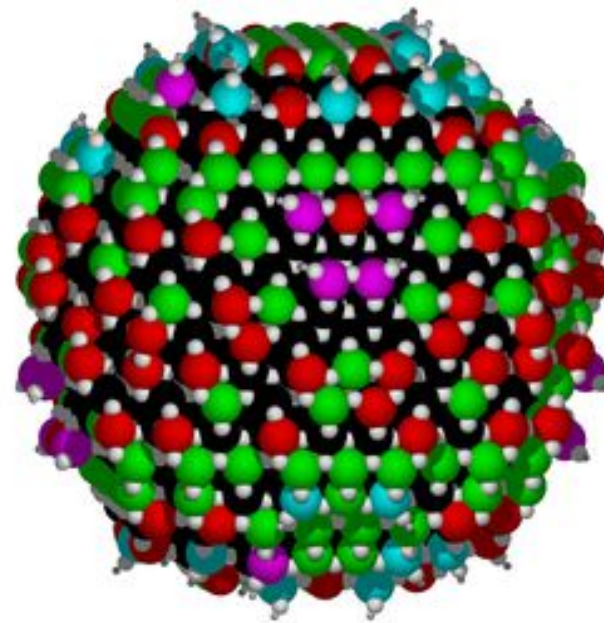

(TiO$_2$)$_{1265}$ · 26H$_2$O

| Diameter = | 1.5 nm | 2.2 nm | 3.0 nm | 4.4 nm |
|---|---|---|---|---|
| Surface-to-Bulk Ratio = | 1.70 | 0.94 | 0.83 | 0.43 |

# 1.5 nm

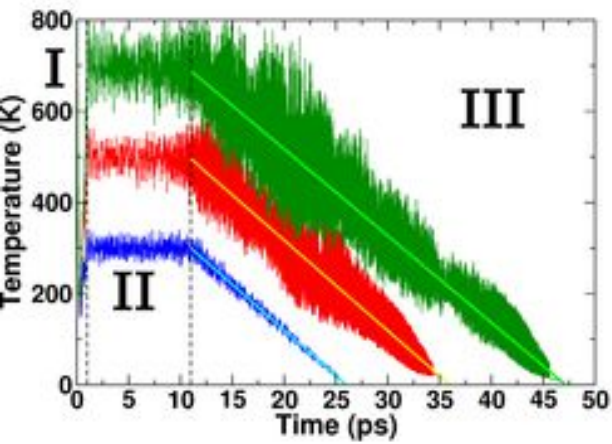

# 2.2 nm

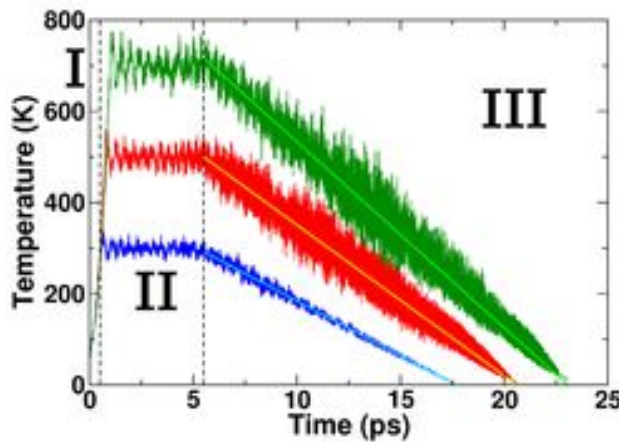

# 3.0 nm

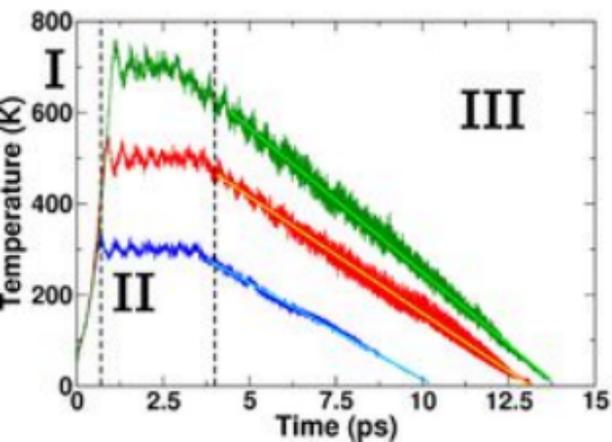

# 4.4 nm

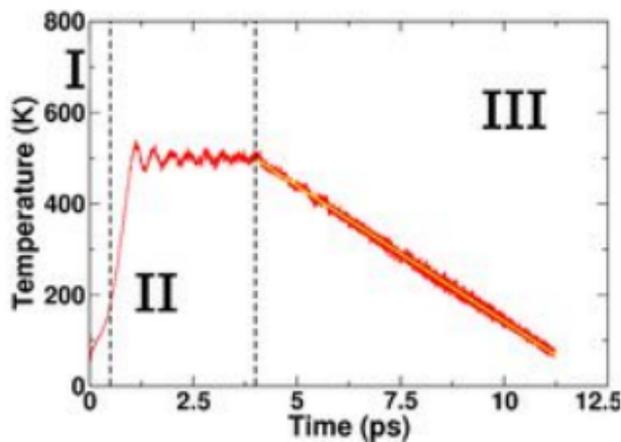

# DFTB
# DFT(B3LYP)

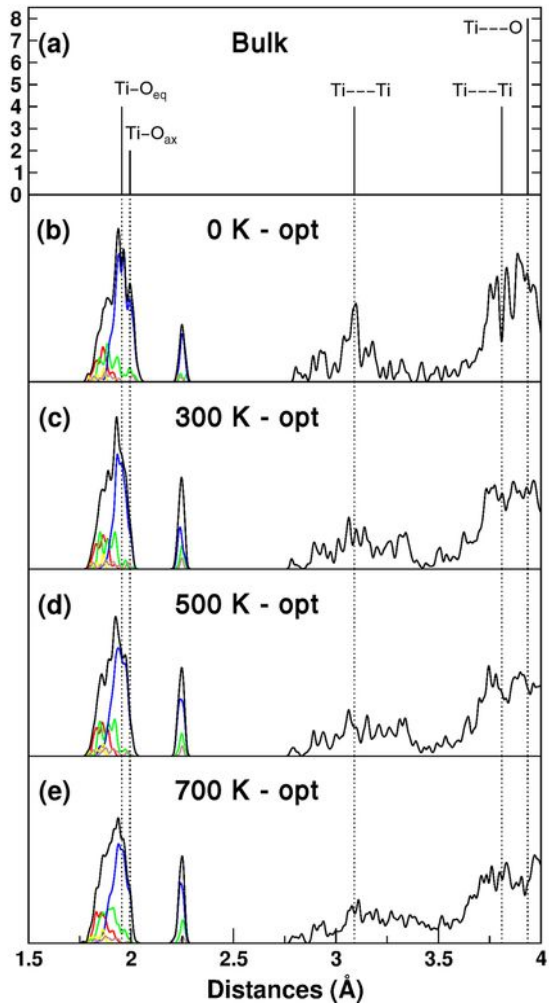
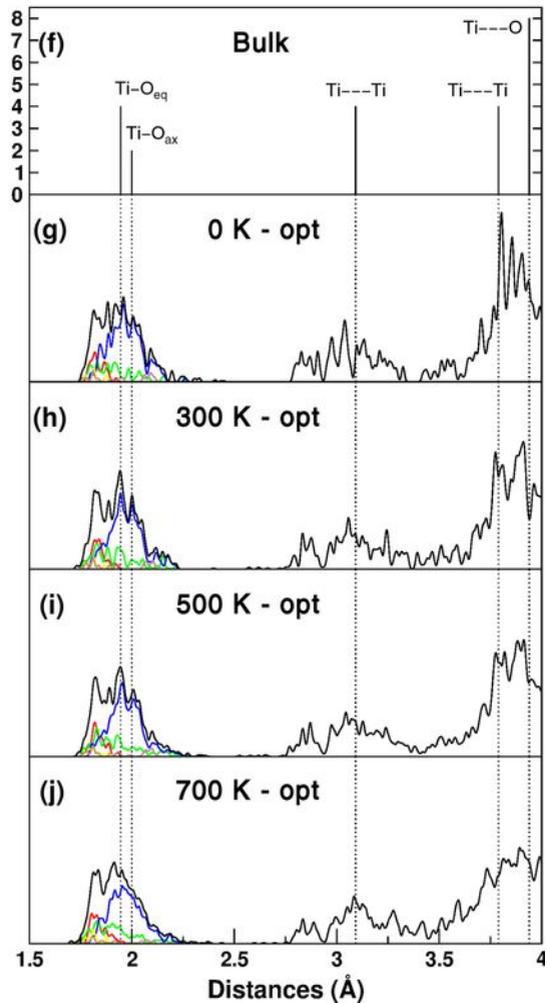

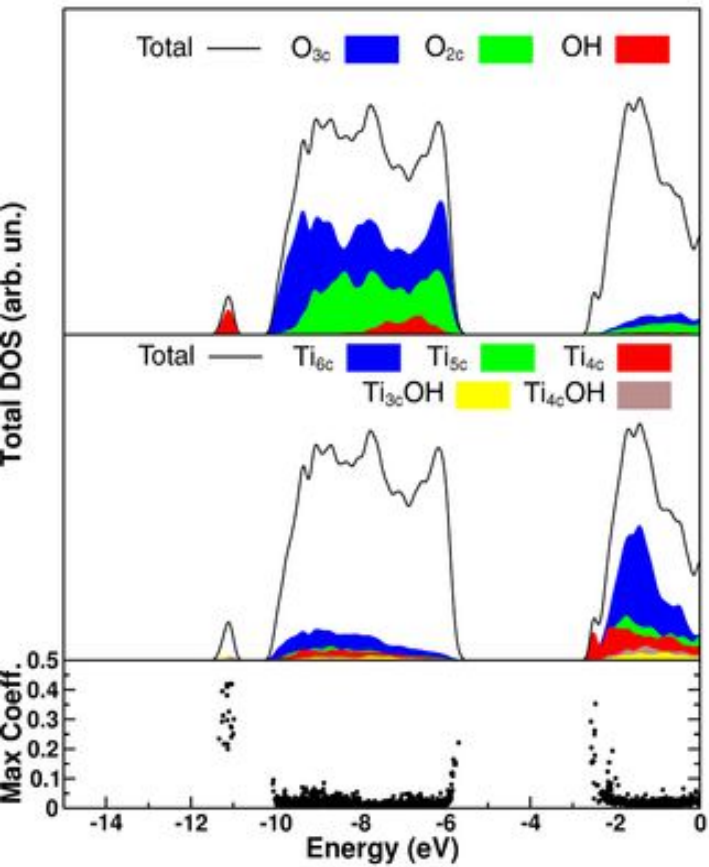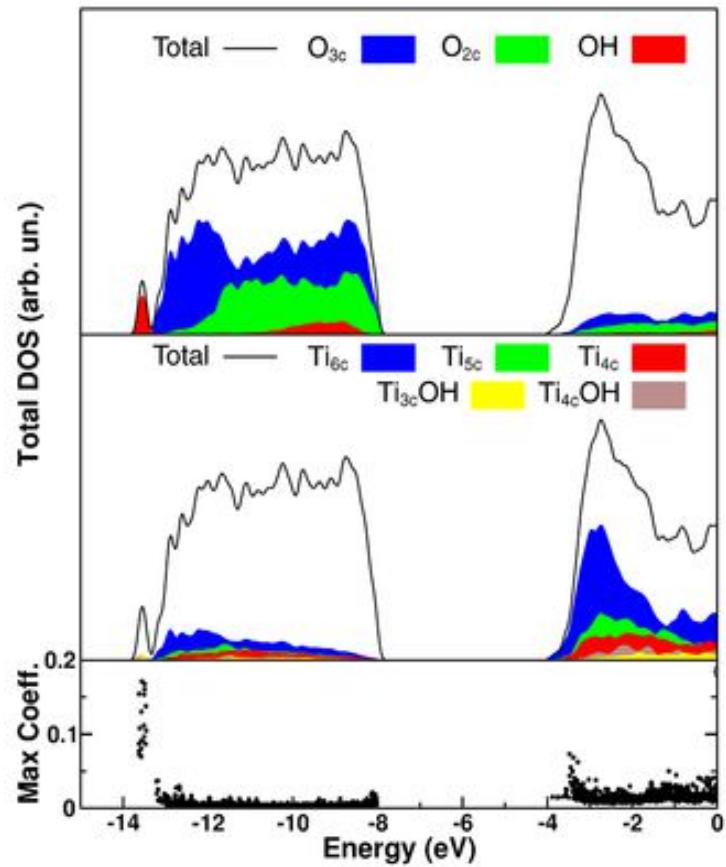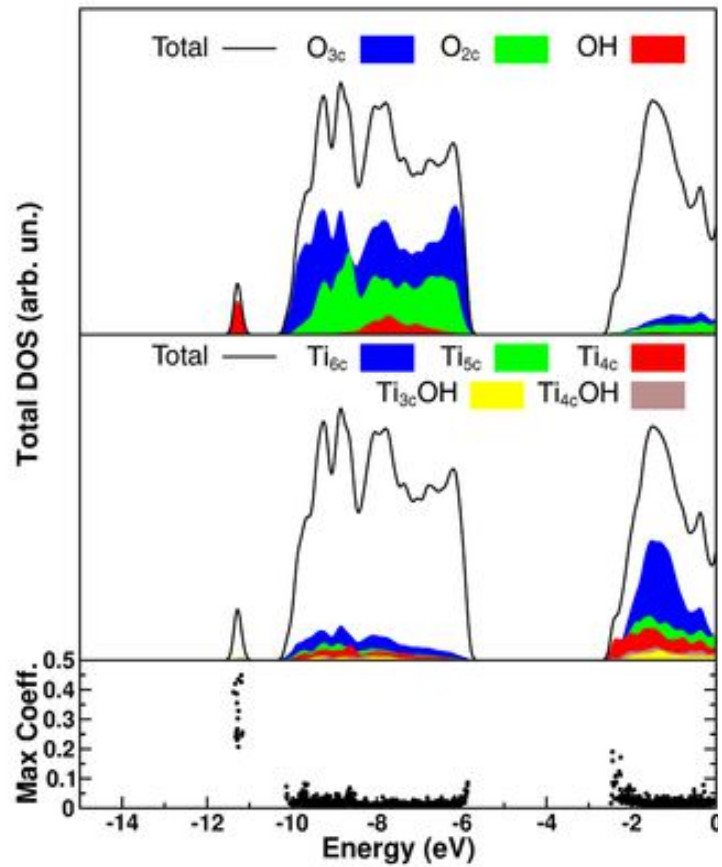

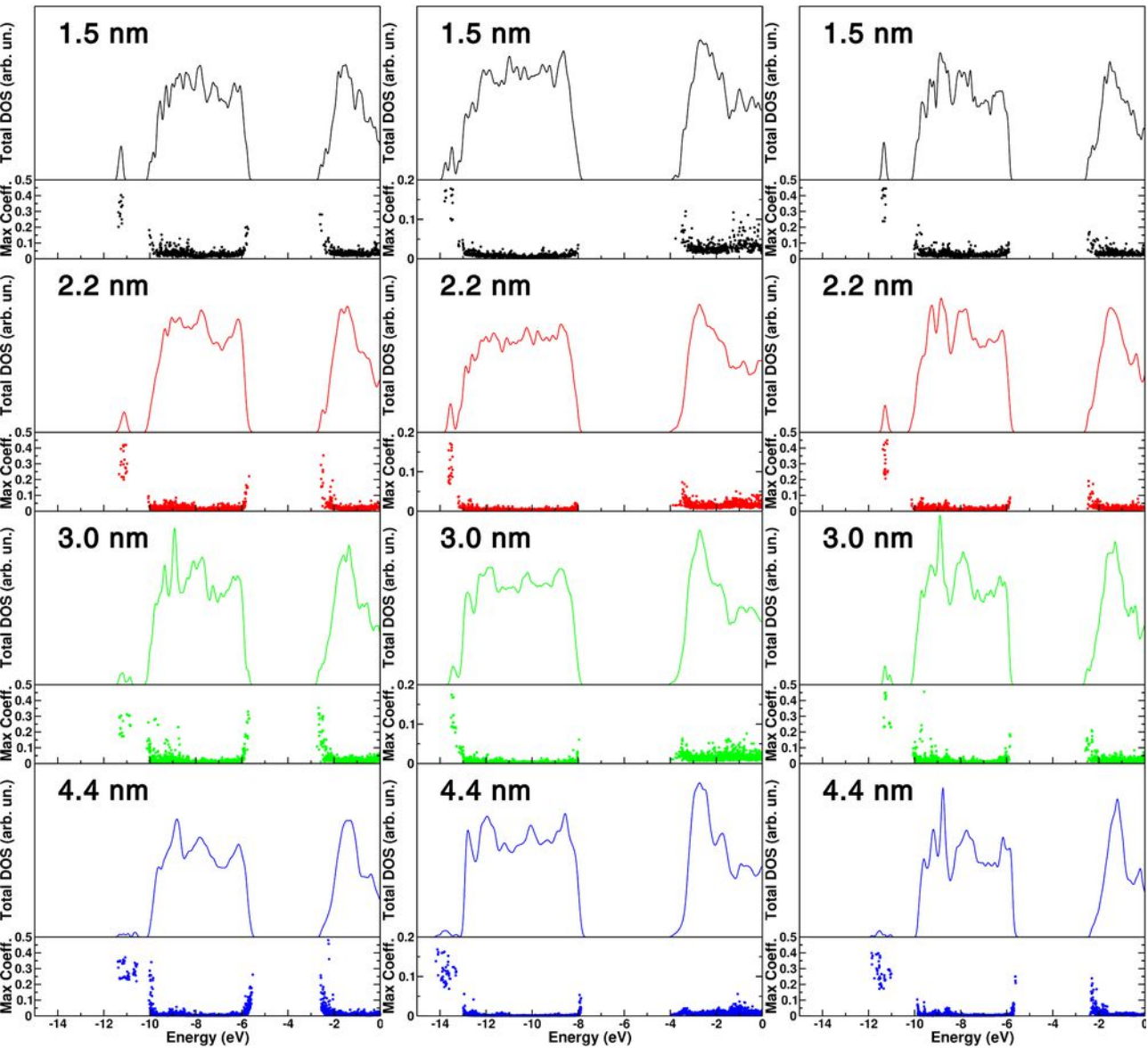